\documentstyle[sprocl]{article}

\input{psfig}
\def\be{\begin{equation}}
\def\ee{\end{equation}}
\def\bea{\begin{eqnarray}}
\def\eea{\end{eqnarray}}

\begin{document}

\title{VIRTUAL COMPTON SCATTERING OFF THE NUCLEON IN THE LINEAR SIGMA MODEL}

\author{ A. METZ \footnote{Talk given at the {\it Workshop on 
 Virtual Compton Scattering}, Clermont-Ferrand, June 1996}, D. DRECHSEL }

\address{Institut f\"ur Kernphysik, Becherweg 45\\
 Johannes Gutenberg--Universit\"at, D--55099 Mainz, Germany}


\maketitle\abstracts{
Virtual Compton scattering off the nucleon has been studied in the
one--loop approximation of the linear sigma model. The three generalized
scalar polarizabilities of the nucleon have been calculated and compared
with the existing theoretical predictions. In particular, we find that
only two of the three scalar polarizabilities are independent
observables.}

\section{Introduction}

At low energies, real Compton scattering (RCS) off the nucleon has been
discussed for quite some time \cite{Klein55} in terms of the electric
$(\alpha)$ and magnetic $(\beta)$ polarizabilities, which depend on the
excitation spectrum of the nucleon. The complementary, purely electromagnetic
reaction of virtual Compton scattering (VCS), $\gamma^{\ast}+N\to\gamma+N$,
promises to provide even more information about the structure of the
nucleon. First, the polarizabilities in VCS become functions of the
four--momentum transfer $Q^{2}$, and second, the virtual photon carries
also a longitudinal polarization.

The formalism of the generalized polarizabilities of the nucleon was
developed by Guichon, Liu and Thomas \cite{Guichon95}. The same authors
gave a first estimate of the generalized polarizabilities of the proton
in the framework of a nonrelativistic constituent quark model \cite{Guichon95}
(CQM), and investigated a more detailed version of the model including
recoil effects \cite{Liu96}. Using an effective lagrangian model (ELM)
Vanderhaeghen \cite{Vanderhaeghen96} determined the $Q^{2}$ behaviour
of $\alpha$ and $\beta$ of the proton. Recently, the slopes of the electric
and magnetic polarizability were predicted by Hemmert et al. \cite{Hemmert96}
in heavy baryon chiral perturbation theory (HBChPT) to order $p^{3}$.
The first experiments \cite{MAMI_prop,CEBAF_prop} to measure the
$Q^{2}$ evolution of the electric polarizability of the proton will soon
provide data.

We have performed a one--loop calculation of the scalar polarizabilities of
both nucleons in the linear sigma model (LSM) \cite{Gell-Mann60} in the
limit of an infinite sigma mass \cite{Metz96}. The sigma model contains
all symmetries relevant for hadron physics at low energies.
The model is Lorentz, gauge and chirally invariant, in particular it
obeys the PCAC relation.

\section{Definitions and Generalized Polarizabilities}

The initial (final) photon of the VCS reaction is characterized by the
polarization vector $\epsilon^{\mu}=(0,\hat{\epsilon})\;
(\epsilon'^{\mu}=(0,\hat{\epsilon}'))$ and the four--momentum
$q^{\mu}=(\omega,\vec{q}\,)
\;(q'^{\mu}=(\omega',\vec{q}^{\hspace{0.1cm}\prime}))$.
We define the momentum transfer as $Q^{2}=-q^{2}=\bar{q}^{2}-\omega^{2}$,
where $\bar{q}=|\vec{q}\,|$ is the absolute value of
the three--momentum of the virtual photon. In the {\it cm} frame,
the Lorentz vector of the incoming (outgoing) nucleon reads
$p^{\mu}=(E,-\vec{q}\,)\;(p'^{\mu}=
(E',-\vec{q}^{\hspace{0.1cm}\prime}))$.
For convenience we introduce the two variables
\begin{eqnarray} \label{gl2_1}
\omega_{0}&=&\omega(\omega'=0)=m_{N}-E=-\frac{\bar{q}^{2}}{2m_{N}}
 +{\mathcal{O}}(\bar{q}^{4})\;,
 \\
 \label{gl2_2}
Q_{0}^{2}&=&Q^{2}(\omega'=0)=2m_{N}\biggl[\sqrt{\bar{q}^{2}+m_{N}^{2}}
 -m_{N}\biggr]\;.
\end{eqnarray}
The expansion in (\ref{gl2_1}) is valid if $\bar{q}$ is small in
comparison to the nucleon mass.

The low energy theorem states \cite{Guichon95,Scherer96} that the leading
order terms of the scattering amplitude, in an expansion in $\omega'$,
are completely determined by the Born amplitude, which depends only on
ground state properties of the nucleon. The excitations of the nucleon
as described by the generalized polarizabilities are of relative order
$\omega'^{2}$.
These polarizabilities are given by the multipoles
$H_{NB}^{(\rho'L',\rho,L)S}(\omega',\bar{q})$ of the non Born amplitude
\cite{Guichon95}. In this notation $\rho(\rho')$ indicates
the type of the initial (final) photon $(\rho=0: \textrm{charge},\;
\rho=1: \textrm{magnetic},\;\rho=2: \textrm{electric})$, and $L(L')$
characterizes its angular momentum. The quantum number $S$ distinguishes
between the no spin--flip $(S=0)$ and the spin--flip $(S=1)$ contributions.
In the Siegert limit, the electromagnetic multipoles have a well--defined
dependence on the momenta of the real or virtual photons. Therefore the
generalized polarizabilities have been defined as \cite{Guichon95}
\begin{eqnarray} \label{gl2_3}
P^{(\rho' L',\rho L)S}(\bar{q})&=&\left[\frac{1}{\omega'^{L}\bar{q}^{L}}
 H_{NB}^{(\rho' L',\rho L)S}(\omega',\bar{q})\right]_{\omega'=0}
 \qquad(\rho,\rho'=0,1)\;, \\
\hat{P}^{(\rho' L',L)S}(\bar{q})&=&\left[\frac{1}{\omega'^{L'}\bar{q}^{L+1}}
 \hat{H}_{NB}^{(\rho' L',L)S}(\omega',\bar{q})\right]_{\omega'=0}\;.
 \nonumber
\end{eqnarray}
The generalized polarizabilities are functions of $\bar{q}$,
or alternatively functions of $Q_{0}^{2}$ (see eq. (\ref{gl2_2})).
The different treatment of the electric multipole for the virtual photon
is due to the fact that it is related to the charge multipole in the
Siegert limit. Therefore, only the difference of these two multipoles
is an independent quantity, the mixed polarizability $\hat{P}$ in
eq. (\ref{gl2_3}).

In the following we will restrict the discussion to the three scalar
polarizabilities $(S=0)$. Two of them generalize the electric
and magnetic polarizability to the virtual photon case,
\begin{equation} \label{gl2_4}
\alpha(Q_{0}^{2})=-\frac{e^{2}}{4\pi}\sqrt{\frac{3}{2}}
 P^{(01,01)0}(Q_{0}^{2})\;,\quad
\beta(Q_{0}^{2})=-\frac{e^{2}}{4\pi}\sqrt{\frac{3}{8}}
 P^{(11,11)0}(Q_{0}^{2})\;,
\end{equation}
the third one is a mixed polarizability, $\hat{P}^{(01,1)0}$. These
polarizabilities define the spin--independent part of the non Born
amplitude to lowest order in $\omega'$,
\begin{eqnarray} \label{gl2_5}
\sqrt{\frac{m_{N}}{E}}T_{NB}^{long}\!\!&=&\!\!\omega'\omega_{0}
 \alpha(Q_{0}^{2})\hat{\epsilon}'^{\star}\cdot\hat{q}
 +{\mathcal{O}}(\omega'^{2})\;,
 \\
\sqrt{\frac{m_{N}}{E}}T_{NB}^{trans}\!\!&=&\!\!\left[\omega'\bar{q}
 \cos\vartheta\beta(Q_{0}^{2})+\omega'\omega_{0}\alpha(Q_{0}^{2})
 -\frac{3e^{2}}{8\pi}\omega'\bar{q}^{2}\hat{P}^{(01,1)0}(Q_{0}^{2})\right]
 \hat{\epsilon}'^{\star}\cdot\hat{\epsilon}\nonumber\\
 & &-\omega'\bar{q}\beta(Q_{0}^{2})\hat{\epsilon}'^{\star}\cdot\hat{q}
    \hat{\epsilon}\cdot\hat{q}'+{\mathcal{O}}(\omega'^{2})
    \vphantom{\frac{1}{1}}\;.\nonumber
\end{eqnarray}

\section{Results}
\begin{figure}
\psfig{figure=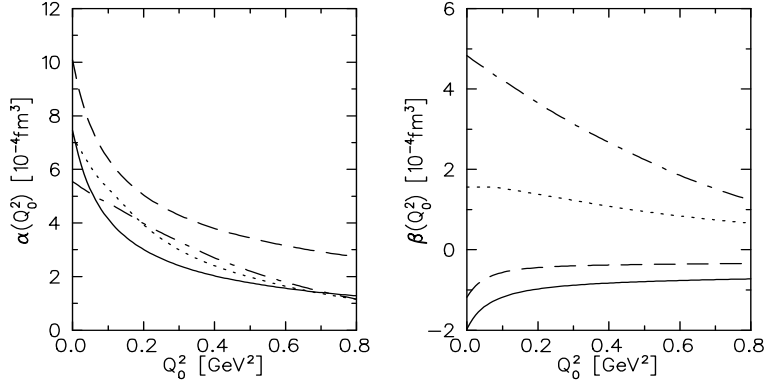,height=5cm,width=10cm}
\caption{The electric and magnetic polarizabilities as function of
 the momentum transfer $Q_{0}^{2}$. Solid line: calculation with the
 LSM for the proton, dashed
 line: LSM result for the neutron, dash--dotted line: CQM
 \protect\cite{Liu96} (proton), dotted line: ELM
 \protect\cite{Vanderhaeghen96} (proton).
 \label{fig1}}
\end{figure}

Using the defining eqs. (\ref{gl2_5}) we have determined the scalar
polarizabilities
for arbitrary $Q_{0}^{2}$. As has been shown in Fig. \ref{fig1}, both
$\alpha$ and $\beta$ are underestimated at the real photon point
$(Q_{0}^{2}=0)$. The corresponding experimental values are, in units of
$10^{-4}\textrm{fm}^{3}$,
$\alpha_{p}^{exp}(0)=12.1\pm1.0,\;\beta_{p}^{exp}(0)=2.1\mp1.0,\;
\alpha_{n}^{exp}(0)=12.6\pm2.5,\;\beta_{n}^{exp}(0)=3.2\mp2.5\;$
\cite{MacGibbon95,Schmiedmayer91}.
In general, we achieve a better description for $\alpha(0)$ than
for $\beta(0)$. The shortcoming of the magnetic polarizability is
due to the neglect of the $\Delta$ resonance in the LSM.

Fig. \ref{fig1} also compares our results with the predictions of the CQM
\cite{Liu96} and the ELM \cite{Vanderhaeghen96}. The various models rely on
quite different pictures of the nucleon. While in the LSM the excitations
of the nucleon are given by pion--nucleon scattering states,
the CQM describes the excitation spectrum by a number of resonances of the
nucleon. The ELM contains both degrees of freedom in the form of effective
lagrangians. As a consequence the three models lead to a different
$Q^{2}$ dependence of the polarizabilities. At low $Q^{2}$, the LSM predicts a
rapid decrease (increase) of the electric (magnetic) polarizability.
In contrast to this the CQM results in a rather smooth decrease of
the polarizabilities close to the real photon point.

In the real photon limit, we are able to derive analytical results.
Restricting ourselves to the proton, the polarizabilities may be expanded
in the mass ratio $\mu=m_{\pi}/m_{N}$,
\begin{eqnarray} \label{gl3_1}
 \alpha_{p}(0)&=&\frac{e^{2}g_{\pi N}^{2}}{192\pi^{3}m_{N}^{3}}
    \biggl[\frac{5\pi}{2\mu}+18\ln\mu+\frac{33}{2}
    +{\cal{O}}(\mu)\biggr]\,, \nonumber\\
 \beta_{p}(0)&=&\frac{e^{2}g_{\pi N}^{2}}{192\pi^{3}m_{N}^{3}}
    \biggl[\frac{\pi}{4\mu}+18\ln\mu+\frac{63}{2}
    +{\cal{O}}(\mu)\biggr]\,, \\
 \frac{e^{2}}{4\pi}\hat{P}_{p}^{(01,1)0}(0)&=&
    \frac{e^{2}g_{\pi N}^{2}}{192\pi^{3}m_{N}^{4}}
    \biggl[-\frac{11\pi}{12\mu}-12\ln\mu-16+{\cal{O}}(\mu)\biggr]\,,
    \nonumber
\end{eqnarray}
where $g_{\pi N}=13.4$ denotes the coupling strength of the pseudoscalar
pion--nucleon coupling. Our results for $\alpha(0)$ and $\beta(0)$ are
in complete agreement with the predictions of a one--loop
calculation in relativistic ChPT \cite{Bernard92.1}. The expansion of the
derivatives takes the form
\begin{eqnarray} \label{gl3_2}
\frac{d}{dQ^{2}}\alpha_{p}(0)\!\!\!\!&=&\!\!\!\!
   \frac{e^{2}g_{\pi N}^{2}}{3840\pi^{3}m_{N}^{5}}
   \biggl[-\frac{7\pi}{\mu^{3}}\!+\!\frac{68}{\mu^{2}}
   \!-\!\frac{319\pi}{4\mu}\!-\!445\ln\mu\!-\!\frac{3655}{12}
   \!+\!{\cal{O}}(\mu)\biggr]\,, \nonumber\\
\frac{d}{dQ^{2}}\beta_{p}(0)\!\!\!\!&=&\!\!\!\!
   \frac{e^{2}g_{\pi N}^{2}}{3840\pi^{3}m_{N}^{5}}
   \biggl[\frac{\pi}{\mu^{3}}\!+\!\frac{16}{\mu^{2}}
   \!-\!\frac{40\pi}{\mu}\!-\!365\ln\mu\!-\!\frac{4505}{12}
   \!+\!{\cal{O}}(\mu)\biggr]\,, \\
\frac{e^{2}}{4\pi}\frac{d}{dQ^{2}}\hat{P}_{p}^{(01,1)0}(0)\!\!\!\!&=&\!\!\!\!
   \frac{e^{2}g_{\pi N}^{2}}{3840\pi^{3}m_{N}^{6}}
   \biggl[\frac{2\pi}{\mu^{3}}\!-\!\frac{28}{\mu^{2}}
   \!+\!\frac{89\pi}{2\mu}\!+\!330\ln\mu\!+\!\frac{920}{3}
   \!+\!{\cal{O}}(\mu)\biggr]\,. \nonumber
\end{eqnarray}
Both the leading $m_{\pi}^{-1}$ term in (\ref{gl3_1}) and the
$m_{\pi}^{-3}$ contribution in (\ref{gl3_2}) for the electric and magnetic
polarizability coincide with the predictions of HBChPT to third order in
the external momenta \cite{Bernard92.2,Hemmert96}. These leading terms
are completely fixed by the one--loop calculation, while the other
terms of the expansion will be modified by higher loops and additional
low energy constants.

\section{Relation between the Scalar Polarizabilities}

Our analytical results in (\ref{gl3_1}) and (\ref{gl3_2})
obey the equations
\begin{eqnarray} \label{gl4_1}
\frac{e^{2}}{4\pi}\hat{P}^{(01,1)0}(0)\!\!&=&\!\!-\frac{1}{3m_{N}}
 \Bigl[\alpha(0)+\beta(0)\Bigr]\;, \\
\frac{e^{2}}{4\pi}\frac{d}{dQ_{0}^{2}}\hat{P}^{(01,1)0}(0)
 \!\!&=&\!\!-\frac{1}{3m_{N}}\Bigl[\frac{d}{dQ_{0}^{2}}\alpha(0)
    +\frac{d}{dQ_{0}^{2}}\beta(0)\Bigr]
    +\frac{1}{12m_{N}^{3}}\Bigl[\alpha(0)+\beta(0)\Bigr]\;, \nonumber
\end{eqnarray}
which are valid to all orders in $\mu$. Numerically we have also established
the more general result
\begin{equation} \label{gl4_2}
\frac{e^{2}}{4\pi}\hat{P}^{(01,1)0}(Q_{0}^{2})
 =\frac{2\omega_{0}}{3\bar{q}^{\hphantom{.}2}}\Bigl[\alpha(Q_{0}^{2})
 +\beta(Q_{0}^{2})\Bigr]\;,
\end{equation}
i.e., the scalar polarizabilities are related for the whole range of
$Q_{0}^{2}$. As a consequence the electric and the mixed polarizability
may be eliminated from the transverse amplitude in eq. (\ref{gl2_5}), i.e.,
the transverse amplitude is completely determined by the magnetic
polarizability $\beta$! This result depends, of course, on the fact
that only terms up to $\omega'$ have been considered in (\ref{gl2_5}).
The relation (\ref{gl4_2}) between the polarizabilities leads to the
even more surprising result
\begin{equation} \label{gl4_3}
 H^{(21,21)0}(\omega',\bar{q})=\frac{4\pi}{e^{2}}\sqrt{\frac{8}{3}}
 \omega'\omega_{0}\beta(Q_{0}^{2})+{\mathcal{O}}(\omega'^{2})\;.
\end{equation}
According to (\ref{gl4_3}) the transverse electric multipole, to linear
order in $\omega'$, is determined by the magnetic polarizability.
Since $\omega_{0}$ vanishes in the static limit, $m_{N}\to\infty$,
the right hand sides of both eq. (\ref{gl4_2}) and (\ref{gl4_3}) vanish
in that limit, and the corresponding observables are recoil effects
to that order.

It can be shown that (\ref{gl4_2}) is not a peculiarity of the
LSM but a model--independent result \cite{Knoechlein96}.
The proof relies on a low energy expansion of the non Born
amplitude. The spin--averaged Compton
tensor $T_{NB}^{\mu\nu}$
$(T_{NB}=\epsilon_{\mu}\epsilon_{\nu}T_{NB}^{\mu\nu})$ has the general
structure
\begin{eqnarray} \label{gl4_4}
T_{NB}^{\mu\nu}&=&\left[q'^{\mu}q^{\nu}-q\cdot q'g^{\mu\nu}\right]f_{1}
 \\
& &+\left[P\cdot q'(P^{\mu}q^{\nu}+q'^{\mu}P^{\nu})
           -q\cdot q'P^{\mu}P^{\nu}-(P\cdot q')^{2}g^{\mu\nu}\right]f_{2}
 \nonumber\\
& &+\left[P\cdot q'q^{2}g^{\mu\nu}-P\cdot q'q^{\mu}q^{\nu}
           -q^{2}q'^{\mu}P^{\nu}+q\cdot q'q^{\mu}P^{\nu}\right]f_{3}\;,
 \nonumber
\end{eqnarray}
where $P=p+p'$ and $f_{i}=f_{i}(Q^{2},q\cdot q',P\cdot q'),\; i=1,2,3$.
Due to charge conjugation symmetry and nucleon crossing the Compton tensor
is even as function of $P$,
\begin{equation} \label{gl4_5}
T_{NB}^{\mu\nu}(q,q',P)=T_{NB}^{\mu\nu}(q,q',-P)\;.
\end{equation}
Accordingly, $f_{3}$ is odd as function of $P\cdot q'$ and therefore at least
linear in $\omega'$. Since the tensor structure in front of $f_{3}$ is also
linear in $\omega'$, we may neglect the contribution
of $f_{3}$ to the scattering amplitude if we are only interested in
terms linear in $\omega'$. Hence, the transverse amplitude to that order
is given by
\begin{eqnarray} \label{gl4_6}
T_{NB}^{trans}&=&\left[-\omega'\bar{q}\cos\vartheta f_{1}(Q_{0}^{2},0,0)
 +\omega'\omega_{0} f_{1}(Q_{0}^{2},0,0)\right]
 \hat{\epsilon}'^{\star}\cdot\hat{\epsilon} \\
 & &+\omega'\bar{q} f_{1}(Q_{0}^{2},0,0)\hat{\epsilon}'^{\star}\cdot\hat{q}
    \hat{\epsilon}\cdot\hat{q}'+{\mathcal{O}}(\omega'^{2})\;. \nonumber
\end{eqnarray}
Comparing this equation with the low energy expansion (\ref{gl2_5})
leads to relation (\ref{gl4_2}) between the scalar polarizabilities.
This result is generally difficult to obtain in phenomenological
calculations, bcause it requires not only Lorentz and gauge symmetry but
also symmetry under charge conjugation and nucleon crossing.

\section*{Acknowledgments}
This work has been supported by the Deutsche Forschungsgemeinschaft (SFB 201).
We would like to thank G. Kn\"ochlein and S. Scherer for many stimulating
discussions. One of us (A.M.) is also indebted to P.A.M. Guichon and
M. Vanderhaeghen for clarifying conversations about the formalism of the
generalized polarizabilities.

\end{document}